\documentclass[10pt]{IEEEtran}
\usepackage{fancyhdr}
\usepackage{cite}
\usepackage{url}
\usepackage{amsmath,amssymb,amsfonts}
\usepackage{graphicx}
\usepackage{textcomp}
\usepackage{xcolor}
\usepackage[noend]{algpseudocode}
\usepackage{algorithm}
\usepackage{algorithmicx}
\usepackage{booktabs}
\def\BibTeX{{\rm B\kern-.05em{\sc i\kern-.025em b}\kern-.08em
    T\kern-.1667em\lower.7ex\hbox{E}\kern-.125emX}}
\usepackage[left=1.43cm,right=1.43cm,top=1.96cm,includefoot,bottom=1 in]{geometry}

\IEEEoverridecommandlockouts\IEEEpubid{\makebox[\columnwidth]{ 978-1-6654-3540-6/22~\copyright~2022 IEEE \hfill} \hspace{\columnsep}\makebox[\columnwidth]{ }}

\begin{document}

\title{Resource Allocation and Resolution Control in the Metaverse with Mobile Augmented Reality}
\author{\IEEEauthorblockN{Peiyuan Si$^1$,
Jun Zhao$^1$, Huimei Han$^{1,2}$,
Kwok-Yan Lam$^1$, Yang Liu$^{1,3}$}
\IEEEauthorblockA{\\$^1$School of Computer Science \& Engineering, Nanyang Technological University, Singapore\\$^2$College of Information Engineering, Zhejiang University of Technology, Hangzhou, Zhejiang, P.R. China\\$^3$School of Information and Communication Engineering, Dalian University of Technology, Dalian, China\\
peiyuan001@e.ntu.edu.sg, \{junzhao, huimei.han, kwokyan.lam\}@ntu.edu.sg, yangliu\_613@dlut.edu.cn\vspace{-20pt}} }

\maketitle
\thispagestyle{fancy}
\pagestyle{fancy}
\lhead{This paper appears in the Proceedings of IEEE Global Communications Conference (GLOBECOM) 2022.\\ Please feel free to contact us for questions or remarks.}
\cfoot{~\\[-30pt]\thepage}

\begin{abstract}
With the development of blockchain and communication techniques, the Metaverse is considered as a promising next-generation Internet paradigm, which enables the connection between reality and the virtual world. The key to rendering a virtual world is to provide users with immersive experiences and virtual avatars, which is based on virtual reality (VR) technology and high data transmission rate. However, current VR devices require intensive computation and communication, and users suffer from high delay while using wireless VR devices. To build the connection between reality and the virtual world with  current technologies, mobile augmented reality (MAR) is a feasible alternative solution due to its cheaper communication and computation cost. This paper proposes an MAR-based connection model for the Metaverse, and proposes a communication resources allocation algorithm based on outer approximation (OA) to achieve the best utility. Simulation results show that our proposed algorithm is able to provide users with basic MAR services for the Metaverse, and outperforms the benchmark greedy algorithm.
\end{abstract}


\section{Introduction}
The development of blockchain and communication techniques motivated intensive interest in the Metaverse, which is considered a next-generation Internet paradigm \cite{xu2021wireless, DusitDynamicResource, cheng2022will}. Attracted by the potential of the Metaverse, many governments and companies around the world are planning and preparing for the upcoming Metaverse era; e.g., South Korea is promoting lessons based on
the Metaverse, Facebook announced that it would become a Metaverse company and renamed itself as Meta, and Tencent has invested in an AR platform called ``Avakin life" \cite{ningSTA2021survey, Taxtonomy2022metaverse, facebook2021critical}.

\textbf{Current challenges.} One of the key issues in the Metaverse is to connect the virtual world and the real world with the support of extend reality (XR), including virtual reality (VR) and augmented reality (AR) \cite{AllOneNeedToKnow, hu2020cellular, Akyildiz2022XR, THz2022can}. Current research on the Metaverse mainly focuses on solving the problem of communication and computation for VR to provide users with immersive experience \cite{zhou2020communication, liu2021learning}. However, mobile VR services require extremely high data rate, which is not easy to achieve even under the context of 5G. Besides, VR users suffer and have to bear with the high weight of current VR devices, which is another problem to be solved.

Mobile augmented reality (MAR) is a possible alternative option for VR, and also an important component of the Metaverse \cite{huang2021proactive, MAR2021survey, MAR_InfoCom2018edge, MAR2017context}. Compared to VR, MAR combines reality and the virtual world rather than creating a fully virtual world, which saves communication and computational cost \cite{siriwardhana2021AR,MARl2018caching, MAR2019edge}. Besides, current AR devices have a significant advantage over VR in weight, making them more comfortable and safer as wearable devices. AR also has its unique advantage in some applications, e.g., navigation, health care, tourism, shopping and education, where the interaction with reality is required \cite{xi2022challenges,Tourism2022mixed, MAR2021Health}.

\textbf{Related work.}
Although AR requires lower data rate than VR, efficient allocation of communication resources is still necessary due to the massive number of users and devices connected to the Metaverse server. To improve the communication resource efficiency and quality of service (QoS), MEC and reinforcement learning (RL) for VR/AR service have attracted much attention \cite{energy2021, sun2019communications, liu2018dare, guo2020adaptive, wang2020joint, RL_VR2020smart, RL_VR2021learning}. Feng \emph{et al.} \cite{RL_VR2020smart} proposed a smart VR transmission mode scheme based on RL to optimize the D2D system throughput and achieve a balance between performance and resource efficiency. Chen \emph{et al.} \cite{energy2021} introduced an RL-based energy-efficient MEC framework for AR services with task offloading and resource allocation to release the burden at the terminal.   A recent work~\cite{Chua2022} applies deep RL to MAR services of the metaverse over 6G wireless networks.
Resolution control is also one of the solutions to improve resource efficiency \cite{resolution_control}.
Higher resolution brings better QoS at the cost of occupying more communication resources while lower resolution improves resource efficiency. Thus, finding the balance between QoS and communication resources such as power and bandwidth is of vital importance to Metaverse MAR service.

In this paper, we propose an MAR-based Metaverse model with resolution control and resource allocation algorithm. In our proposed resource allocation optimization problem, both QoS and power consumption are included in the utility function for the balance between user experience and the energy efficiency of the service provider. To solve the mixed-integer nonlinear programming (MINLP) optimization problem, we propose a resource allocation algorithm based on outer approximation (OA), which guarantees global optimum. Simulation results show that our algorithm is able to maximize the utility with given communication resources, and outperforms the benchmark greedy algorithm.

\textbf{Contributions.} The contributions of this paper are as follows:
\begin{itemize}
\item A MAR-based Metaverse model is proposed for applications where interaction with reality is required.
\item A resolution control and resource allocation optimization problem for Metaverse MAR services is proposed and solved by an OA algorithm.
\item Simulation of the proposed algorithm is implemented with the comparison to a benchmark greedy algorithm under various parameter settings. The results show the advantage of our proposed algorithm and the feasibility of the proposed Metaverse MAR service model.
\end{itemize}

The rest of this paper is organized as follows. Section II introduces the proposed system model. The problem formulation and solution are presented in Section III and Section IV respectively. Section V shows the simulation results and the corresponding explanation. The conclusion of this paper is discussed in Section VI.

\section{System Model and Problem Formulation}
\begin{figure}[tbp]
  \centering
  \includegraphics[width=0.9\linewidth]{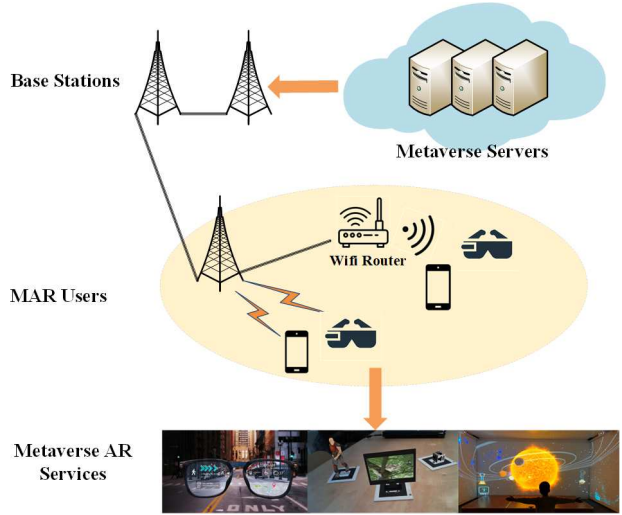}
  \caption{System model of the Metaverse with mobile augmented reality.}
  \label{fig:System_Model}
\end{figure}
In our proposed Metaverse MAR service model, we consider MAR service for multiple users supported by a base station in a particular area, which is connected to Metaverse servers. As shown in Fig. \ref{fig:System_Model}, MAR users can get Metaverse services with data synchronized to their avatars in the Metaverse while interacting with the real world with the support of the base station, which controls the MAR service quality by switching the resolution, e.g., 480P, 720P, 1080P. Due to the limit of total transmission power, the base station needs to dynamically allocate power for users and control the resolution of MAR service according to the channel condition to achieve the best utility.

We assmue that the base station provides $N$ users with MAR service through $N$ channels.
With fixed bandwidth and limited transmit power, the base station allocates transmit power to $N$ users, and the transmission rate for user $n$ is given by

\begin{align}
{{r}_{n}}=B{{\log }_{2}}\left( 1+\frac{{{p}_{n}}{{g}_{n}}}{{{\sigma }^{2}}} \right),
\end{align}
where $B$ and $p_n $ denotes the bandwidth and transmit power for user $n$ respectively, and $\sigma^2$ denotes the power of Gaussian noise. $g_n$ denotes the channel gain between base station and user $n$.  Let $\boldsymbol{p} : = [p_1, p_2, \ldots, p_N]$. The base station is able to provide users with different resolutions of MAR service according to the channel condition. The required transmission rate for real time MAR service is denoted by
\begin{align}
{{C}_{n}}\in \{{{C}_{1}},{{C}_{2}},{{C}_{3}}\},
\end{align}
where ${{C}_{1}},{{C}_{2}},{{C}_{3}}$ denote the required transmission rate of 360P, 720P and 1080P MAR service respectively. The base station can provide each user with only one type of service, and the selection indicator is given by
\begin{align}
\boldsymbol{I} =\{{{I}_{n,1}},{{I}_{n,2}},{{I}_{n,3}}|n\in [N]\},
\end{align}
where ${{I}_{n,1}},{{I}_{n,2}},{{I}_{n,3}}\in \{0,1\}$, $\sum_{i=1}^{3}{{I}_{n,i}}=1$ and $[N]:=\{1,2,..,N\}$.
To evaluate the quality of service (QoS) for users with given service selection, we introduce a QoS factor $Q_n$, which is given by
\begin{align}
{{Q}_{n}}={{\left( \frac{{{C}_{n}}}{{{R}_{th}}} \right)}^{\gamma }}=\sum\limits_{i=1}^{3}{{{I}_{n,i}}{{\left( \frac{{{C}_{i}}}{{{R}_{th}}} \right)}^{\gamma }}},
\end{align}
where $R_{th}$ denotes the constant reference service quality, and $\gamma$ denotes the exponential factor of QoS. The best QoS is given by
\begin{align}
{{Q}_{0}}={{\left( \frac{{{C}_{3}}}{{{R}_{th}}} \right)}^{\gamma}},
\end{align}

QoS only evaluates the experience of individual user, which does not fully represent the utility of the resource allocation scheme. If the base station provides more users with higher QoS under given bandwidth resource, more power will be occupied to reach the required transmission rate. Thus, we take the total power consumption into consideration for the utility function, which is given by
\begin{align}
U=(1-\lambda -\mu ){{\eta }_{q}}\sum\limits_{n=1}^{N}{{{Q}_{n}}}-\lambda {{\eta }_{p}}\sum\limits_{n=1}^{N}{{{p}_{n}}}+\mu {{\eta }_{r}}\sum\limits_{n=1}^{N}{\Delta {{r}_{n}}},
\end{align}
\begin{align}
\Delta {{r}_{n}}=\sum\limits_{i=1}^{3}{{{C}_{i}}I_{n,i}^{(0)}}-B{{\log }_{2}}\left( 1+\frac{{{p}_{n}}{{g}_{n}}}{{{\sigma }^{2}}} \right),
\end{align}
where ${{\eta }_{q}}={{\left( N{{Q}_{0}} \right)}^{-1}}$, ${{\eta }_{r}}={{{{C}_{2}}}^{-1}}$ and ${{\eta }_{p}}={{P}^{-1}}$ are normalization factors, $\lambda$, $\mu$ and $1-\lambda-\mu$ denote the levels of concern for energy consumption, redundancy of transmission rate and QoS respectively. We consider the redundancy of transmission rate as a positive factor because it provides robustness against the turbulence of channel. To maximize the utility of the resource allocation scheme under given total transmission power limit, the optimization problem is given by

\begin{align}
  & P1:\underset{\boldsymbol{I},\boldsymbol{p}}{\mathop{\min }}\,\text{ }\lambda {{\eta }_{p}}\sum\limits_{n=1}^{N}{{{p}_{n}}-\mu {{\eta }_{r}}\sum\limits_{n=1}^{N}{\Delta {{r}_{n}}}-}(1-\lambda -\mu ){{\eta }_{q}}\sum\limits_{n=1}^{N}{{{Q}_{n}}} \nonumber\\
 & subject\text{ }to \nonumber\\
 & Con1:\sum\limits_{i=1}^{3}{{{C}_{i}}I_{n,i}^{(0)}}-B{{\log }_{2}}\left( 1+\frac{{{p}_{n}}{{g}_{n}}}{{{\sigma }^{2}}} \right)\le 0,\forall n\in [N], \nonumber\\
 & Con2:\sum\limits_{n=1}^{N}{{{p}_{n}}}\le P, \nonumber\\
 & Con3:\sum\limits_{i=1}^{3}{I_{n,i}^{{}}}=1,\forall n\in [N],I_{n,i}^{{}}\in \{0,1\},
\end{align}
where constraint $Con1$ indicates that data transmission rate should meet the requirement of the selected service quality. Constraint $Con2$ denotes the total power constraint, and constraint $Con3$ indicates that the base station can provide each user with only one type of service.
Problem $P1$ is a mixed-integer nonlinear programming (MINLP) problem, which can be solved by outer approximating (OA) method. The solution of $P1$ is presented in Section IV.

\section{Problem Solution}
The MINLP problem $P1$ can be solved by OA method in different steps, which are explained in the subsections below.

\subsection{Solve the Nonlinear Programming Problem with Given Integer Variables}
At the beginning of the OA algorithm, i.e., the first iteration, we give initial values to the integer variables matrix $\boldsymbol{I}$ in the feasible region, which are denoted as $\boldsymbol{I}^{(0)}$. Then substitute $\boldsymbol{I}^{(0)}$ into $P1$ to formulate the nonlinear programming problem $P1.1$, which is given by
\begin{align}
  & P1.1:\underset{\boldsymbol{p}}{\mathop{\min }}\,\text{ }\lambda {{\eta }_{p}}\sum\limits_{n=1}^{N}{{{p}_{n}}-\mu {{\eta }_{r}}\sum\limits_{n=1}^{N}{\Delta {{r}_{n}}}}-{{A}_{1}} \nonumber\\
 & subject\text{ }to \nonumber\\
 & Con4:B{{\log }_{2}}\left( 1+\frac{{{p}_{n}}{{g}_{n}}}{{{\sigma }^{2}}} \right)-\sum\limits_{i=1}^{3}{{{C}_{i}}I_{n,i}^{(0)}}\ge 0,\forall n\in [N], \nonumber\\
 & Con5:P-\sum\limits_{n=1}^{N}{{{p}_{n}}}\ge 0,
\end{align}
where the constant $A_1$ is given by
\begin{align}
{{A}_{1}}=\text{ }(1-\lambda -\mu ){{\eta }_{q}}\sum\limits_{n=1}^{N}{\sum\limits_{i=1}^{3}{I_{n,i}^{(0)}{{\left( \frac{{{C}_{i}}}{{{R}_{th}}} \right)}^{2}}}}.
\end{align}

With given integer variables, problem $P1$ is transformed into nonlinear programming (NLP) $P1.1$, which can be solved by interior point method, whose worst case iteration complexity in this paper is ${\mathrm O}\left( \sqrt{n}\log n\log \frac{n}{\epsilon } \right)$ \cite{Complexity_Interior}.

We introduce an auxiliary variable $s_n$ to reformulate the problem as
\begin{align}
  & P1.2:\underset{\boldsymbol{p}}{\mathop{\min }}\,\text{ }\lambda {{\eta }_{p}}\sum\limits_{n=1}^{N}{{{p}_{n}}-\mu {{\eta }_{r}}\sum\limits_{n=1}^{N}{\Delta {{r}_{n}}}}-{{A}_{1}} \nonumber\\
 & subject\text{ }to \nonumber\\
 & Con4:B{{\log }_{2}}\left( 1+\frac{{{p}_{n}}{{g}_{n}}}{{{\sigma }^{2}}} \right)-\sum\limits_{i=1}^{3}{{{C}_{i}}I_{n,i}^{(0)}}={{s}_{n}},\forall n\in [N], \nonumber\\
 & Con5:P-\sum\limits_{n=1}^{N}{{{p}_{n}}}\ge 0, \nonumber \\
 & Con6:{{s}_{n}}\ge 0,\forall n\in [N].
\end{align}

The barrier function is given by
\begin{align}
J(p,\tau )&=\lambda {{\eta }_{p}}\sum\limits_{n=1}^{N}{{{p}_{n}}-\mu {{\eta }_{r}}\sum\limits_{n=1}^{N}{\Delta {{r}_{n}}}}-{{A}_{1}}\nonumber\\
& \quad -\tau \left( \log \left( {{s}_{n}} \right)+\log \left( P-\sum\limits_{n=1}^{N}{{{p}_{n}}} \right) \right),
\end{align}
where $\tau$ is a positive barrier parameter. The perturbed KKT conditions for $n\in[N]$ are given by \cite{InteriorPoint}
\begin{align}
    &\nabla J(p,\tau )-\Delta {{r}_{n}}{{z}_{n}}=0, \\
    &\Delta {{r}_{n}}-{{s}_{n}}=0,\\
    &s_{n}z_{n} =\tau,\\
    &s_n\ge 0,z_n\ge 0,
\end{align}
where $z_n,n\in[N]$ are Lagrange multiplier-inspired dual variables.
The interior point method starts with a feasible point which satisfies the perturbed KKT conditions with small $\tau$, and continue to find smaller $\tau$. As the value of $\tau$ approaches zero, the solution is expected to converge to a point which satisfies the KKT point of problem $P1.1$.

With the obtained optimal power allocation, the solution of $P1.1$ at $t$-th iteration is denoted as $z_{ub}^{(t)}\left( \boldsymbol{p}_{ub}^{(t)} \right)$, which is a lower bound of the global optimum, where $\boldsymbol{p}_{ub}^{(t)}$ denotes the optimal solution at the $t$-th iteration.

\subsection{Solve the Mixed-integer Linear Programming Problem}
After obtaining the optimal solution for $P1.2$, we formulate the first-order Taylor expansion of the original problem $P1$ at point $\boldsymbol{p}_{ub}^{(t)}$ as
\begin{align}
  & P1.3:\underset{\boldsymbol{I,p}}{\mathop{\min }}\,\text{ }\lambda {{\eta }_{p}}\sum\limits_{n=1}^{N}{{{p}_{n}}}-\mu {{\eta }_{r}}\sum\limits_{n=1}^{N}{\left( {{A}_{2}}-\sum\limits_{i=1}^{3}{{{C}_{i}}{{I}_{n,i}}} \right)} \nonumber\\
 &~~~~~~~~~~~~~~-(1-\lambda -\mu ){{\eta }_{q}}\sum\limits_{n=1}^{N}{{{Q}_{n}}} \nonumber\\
 & subject\text{ }to \nonumber\\
 & Con7:{{A}_{2}}-\sum\limits_{i=1}^{3}{{{C}_{i}}{{I}_{n,i}}}\le 0,\forall n\in [N], \nonumber\\
 & Con2,Con3,
\end{align}

where
\begin{align}
{{A}_{2}}=B{{\log }_{2}}\left( 1+\frac{p_{n}^{(t)}{{g}_{n}}}{{{\sigma }^{2}}} \right)+\frac{B{{g}_{n}}}{\ln 2\left( {{\sigma }^{2}}+p_{n}^{(t)}{{g}_{n}} \right)}\left( {{p}_{n}}-p_{n}^{(t)} \right).
\end{align}

    Problem $P1.3$ is a mixed-integer linear programming program (MILP), which can be solved by existing MILP solvers with branch-and-bound algorithm. As MILP problems are NP-hard, there are no polynomial complexity algorithms, and thus the complexity in the worst case is still exponential. However, the branch-and-bound algorithm reduces the average complexity compared to  brute force search \cite{Complexity_Branch_Bound}. The precise estimation of branch-and-bound algorithm complexity requires the probability that {a node in the branch-and-bound tree} is optimal, which is hard to obtain in the optimization problem of this paper. To analyze the complexity of this algorithm, we implemented running time experiments which show acceptable results that the algorithm is able to converge within 30 seconds with CPU i7-9750H and 16GB memory.

The optimal solution of problem $P1.3$ at the $t$-th iteration is denoted as $z_{lb}^{(t)}\left(\boldsymbol{I}_{lb}^{(t)}, \boldsymbol{p}_{lb}^{(t)} \right)$, which is a lower bound of the global optimum.

\subsection{Compare the Gap Between $z_{lb}^{(t)}$ and $z_{ub}^{(t)}$}
As the lower bound and upper bound of global optimum are obtained, the gap between them is given by
\begin{align}
G=\left| z_{ub}^{(t)}-z_{lb}^{(t)} \right|.
\end{align}

The algorithm is considered to be converged when $G\le \epsilon$, where $\epsilon$ is a given precision factor. If $G\ge \epsilon$, substitute the integer variables $\boldsymbol{I}_{lb}^{(t)}$ into the original problem $P1$ for the next iteration.
The overall algorithm is shown in Algorithm 1.

\begin{algorithm}[htb]
\caption{Outer Approximating Algorithm for Communication Resource Allocation in the Metaverse} 
{\bf Initialize:} $t=1$\\
{\bf Input:} 
${\textbf{I}^{(0)}}$\\
\textbf{Step 1.1:} Substitute ${\boldsymbol{I}^{(t-1)}}$ into $P1$ to formulate nonlinear programming (NLP) problem $P1.1$\\
\textbf{Step 1.2:} Solve $P1.1$ by interior method to get lower bound $z_{ub}^{(t)}\left( \boldsymbol{p}_{ub}^{(t)} \right)$ of the global optimum\\
\textbf{Step 2.1:} Formulate the MILP problem $P1.3$ with the first-order Taylor expansion at point $\boldsymbol{p}_{ub}^{(t)}$\\
\textbf{Step 2.2:} Solve $P1.3$ by MILP solvers to get $z_{lb}^{(t)}\left( I_{lb}^{(t)},p_{lb}^{(t)} \right)$\\
\textbf{Step 3:} \textbf{If} $G=\left| z_{ub}^{(t)}-z_{lb}^{(t)} \right|\le \epsilon$,
{\textbf{Output:}} 
$\{{\boldsymbol{I}^{(t)}},{{\boldsymbol{p}}^{(t)}}\}$, \textbf{Else} Go to \textbf{Step 1.1}
\end{algorithm}

\section{Simulation Results}
In this section, we present the simulation results of the proposed OA-based resource allocation algorithm and the comparison with a benchmark greedy algorithm. In the simulation, the setting of constant parameters is given in Table~\ref{tableParameter}, and the benchmark greedy algorithm is given in Algorithm 2.
The channel gain in simulation is set as
\begin{align}
{{g}_{n}}&=20{{\log }_{10}}({{d}_{n}})+20{{\log }_{10}}(f)-147.55, \\
{{d}_{n}}&={{D}_{f}}{{d}_{0}}(n),
\end{align}
where ${{d}_{0}}(n)$ denotes the reference distance between base station and user $n$, and ${{D}_{f}}$ denotes the distance multiply factor. The default value of $\mu$ is set to 0.1.

\begin{table}[tbp]
\caption{Constant Parameter Setting} \label{tableParameter}
\begin{center}
\begin{tabular}{c c}
\toprule[1pt]
 Parameter and Physical Meaning            & Value               \\ \hline
  Exponential factor of QoS ($\gamma$)     & $2$                 \\
  Number of users ($N$)                    & $5$                 \\
  Precision factor ($\epsilon$)            & $10^{-3}$           \\
  Bandwidth ($B$)                          & $5$MHz              \\
  Frequency ($f$)                          & $28$GHz (i.e., 5G spectrum)             \\
  Reference service quality ($R_{th}$)     & $0.77$Mbps          \\
  Power of Gaussian noise ($\sigma^2$)     & $5\times {{10}^{-8}}$W \\
  Reference distance ($d_0$)               & $\left(5,4,3,2,1\right)$(m)           \\
  Required transmission rates $\left(C_1,C_2,C_3\right)$  & $\left(0.77, 1.92, 3.84\right)$(Mbps) \\ \bottomrule[1pt]
\end{tabular}
\end{center}
\end{table}

\begin{algorithm}[tbp]
\caption{Benchmark Greedy Algorithm} 
{\bf Initialize:} $P_{rest} = P$, $U_{max} = -
\infty$\\
\textbf{For} $n \le N$\\
~~~~\textbf{For} $i \le 3$ \\
~~~~$I_{n,i}^{(t)} = 1$, $p_n^{(t)}$ = minimum required power\\
~~~~Calculate the utility $U$ and remaining power $P_{rest}$ under current selection \\
~~~~\textbf{If} $U>U_{max}$ and $P_{rest}\ge 0$\\
~~~~Update $U_{max} = U$, $I_{n,i} = I_{n,i}^{(t)}$, $p_n = p_n^{(t)}$\\
{\textbf{Output:} $U_{max}$, $I_{n,i}$, $p_n$} 
\end{algorithm}

\begin{figure}[tbp]
 \hspace{-12pt}
  \includegraphics[width=1.1\linewidth]{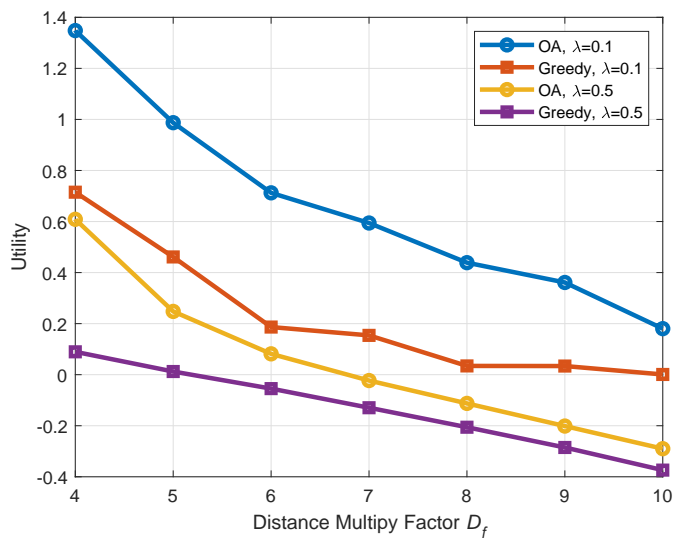}
  \caption{Utility comparison between two algorithms with different $D_f$ and $\lambda$.}
  \label{fig:change_df}
\end{figure}
Fig. \ref{fig:change_df} shows the utility comparison between two algorithms with different $D_f$ and $\lambda$. We can find that the utility of both algorithms decrease with increasing user-to-BS distance because the base station needs more power to provide the same service under smaller channel power gain. The OA algorithm outperforms the greedy algorithm in two different cases, i.e., $\lambda=0.1$ and $\lambda=0.5$, because it ensures global optimum while the greedy algorithm only obtains a local optimum.

\begin{figure}[tbp]
 \hspace{-12pt}
  \includegraphics[width=1.1\linewidth]{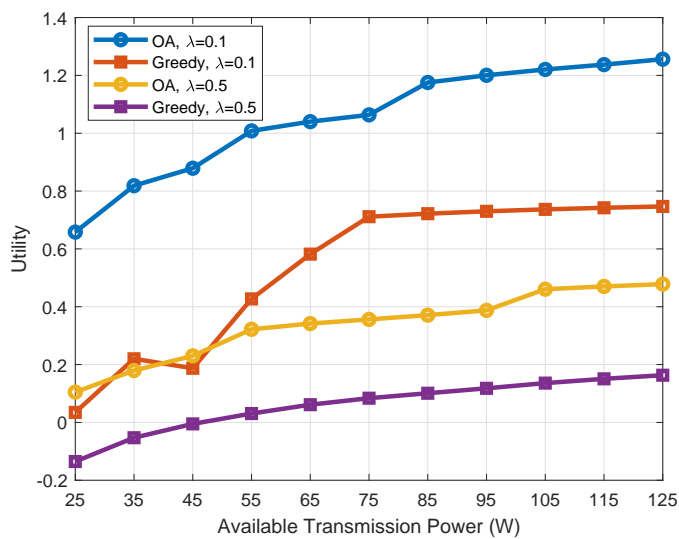}
  \caption{Utility comparison between two algorithms with different $P$ and $\lambda$.}
  \label{fig:change_P}
\end{figure}
Fig. \ref{fig:change_P} shows the utility comparison between two algorithms with different $P$ and $\lambda$. The utilities of both algorithms increase as the total available transmission power $P$ increases because less power is required for the same service, and the base station is able to provide users with better services and higher QoS. Fig. \ref{fig:change_P} also shows the greedy algorithm's ability to find the local optimum and the advantage of OA algorithm in more general cases due to the guarantee of the global optimum.

\begin{figure}[tbp]\hspace{-5pt}
  \includegraphics[width=1.1\linewidth]{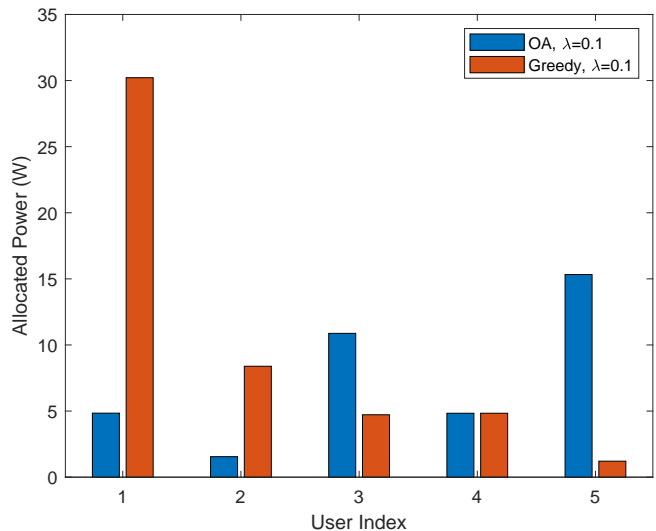}
  \caption{Power allocation with $\lambda=\mu=0.1$, $D_f=5$, $P=50$W.}
  \label{fig:Power_Allocation}
\end{figure}
\begin{figure}[tbp]
  \includegraphics[width=1.1\linewidth]{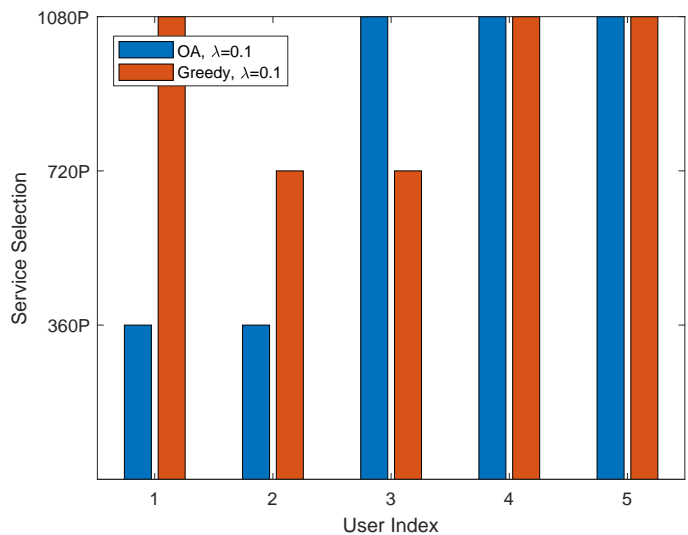}
  \caption{Service selection with $\lambda=\mu=0.1$, $D_f=5$, $P=50$W.}
  \label{fig:Selection}
\end{figure}
Fig. \ref{fig:Power_Allocation} and Fig. \ref{fig:Selection} show the power allocation and service selection under given $D_f=5$, $\lambda=0.1$ and $P=50$W. The powers allocated by the greedy algorithm to users 1, 2 and 3 decrease as the index increase because the distances between the base station and users are $d_{3}<d_2<d_1$.
As the distance decreases, the base station needs less power for basic 360P service. Due to the relatively poor channel condition of users 1, 2 and 3, it is expensive to improve the QoS of these three users. The OA algorithm results in a different policy for power allocation due to its consideration of robustness against channel turbulence, which is also shown in \ref{fig:Selection}. Although user 1 and user 2 are assigned with same service quality, user 1 requires more power due to the longer distance to the base station, and it is the same case for user 3 and user 4. User 5 is assigned with large power because the robustness against channel turbulence contributes to the utility function.

\begin{figure}[tbp]
 \hspace{-12pt}
  \includegraphics[width=1.1\linewidth]{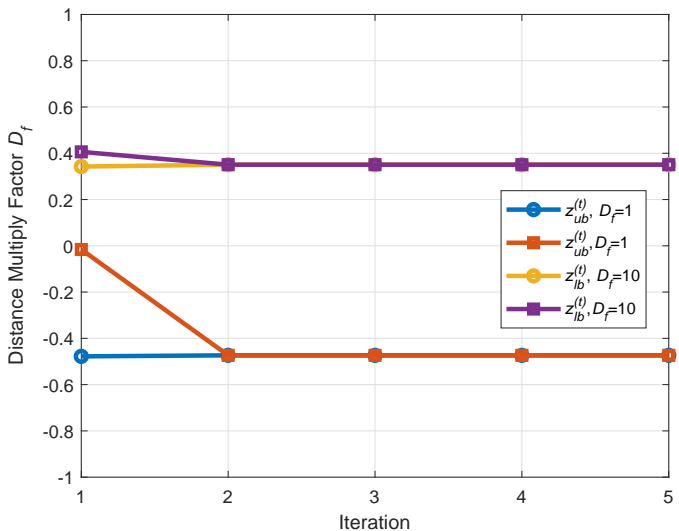}
  \caption{Convergence of OA algorithm.} \vspace{-5pt}
  \label{fig:OA_Convergence}
\end{figure}

Fig. \ref{fig:OA_Convergence} shows the convergence of the proposed OA algorithm for resource allocation under two different sets of parameters. We can find that the upper bound $z_{ub}^{(t)}$ and lower bound $z_{lb}^{(t)}$ converge within 5 iterations regardless of the parameter setting. In the simulation, we set the convergence threshold $\epsilon=10^{-3}$ to limit the gap between $z_{ub}^{(t)}$ and $z_{lb}^{(t)}$, but in most cases the gap becomes zero after several iterations, which further guarantees the convergence.

\begin{figure}[tbp]
 \hspace{-5pt}
  \includegraphics[width=1.1\linewidth]{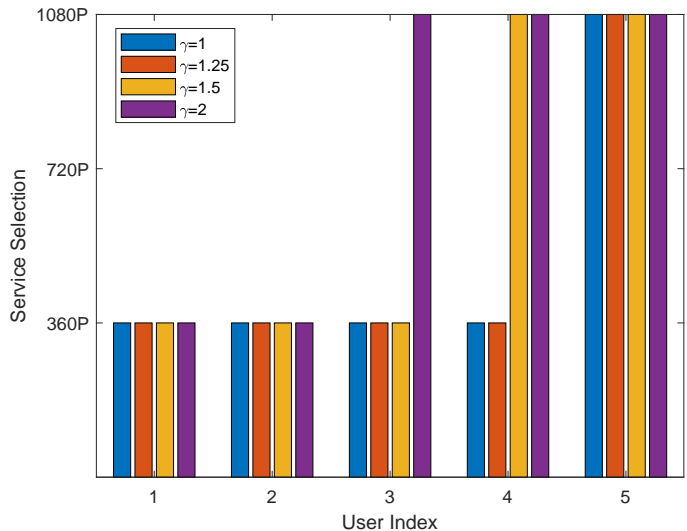}
  \caption{Influence of $\gamma$ on service selection.}\vspace{-5pt}
  \label{fig:OA_gamma}
\end{figure}

Fig. \ref{fig:OA_gamma} shows the influence of $\gamma$ on service selection under $D_f=5$, $P=50$W, $\lambda=0.1$, where the base station takes the power consumption as the main concern, and the power resource is adequate to provide all users with the best service. In this case, the exponential factor of QoS has a great influence on the policy decisions of the base station. With larger $\gamma$, the base station gets more benefit in utility through improving the QoS, and becomes more willing to provide the users with better service even if it gets more penalty from energy consumption. Thus, setting $\gamma$ properly is critical to the balance between power consumption and QoS. {As $\gamma$ increases, the base station tends to improve QoS of the users with larger index first because they are closer to the base station than those with smaller index, which requires less power to achieve the same QoS.}
\section{Conclusion}
In this paper, we proposed a MAR-based Metaverse model for applications that require interaction with the real world. In order to maximize the utility of the base station, we formulate a resolution control and resource allocation optimization problem, and solve the MINLP problem with OA algorithm. The simulation results indicate that our proposed resource allocation algorithm outperforms the benchmark greedy algorithm due the guarantee of global optimum. Our work also shows the feasibility of MAR to provide basic Metaverse service and the advantage of our proposed algorithm over benchmark greedy algorithm.

\section*{Acknowledgement}

This research/project is supported by the National Research Foundation, Singapore under its Strategic Capability Research Centres Funding Initiative. Any opinions, findings and conclusions or recommendations expressed in this material are those of the author(s) and do not reflect the views of National Research Foundation, Singapore. This research/project is supported by the Ministry of Education, Singapore, under its Grant Tier 1 RG97/20, and Grant Tier 1 RG24/20; in part by the NTU-Wallenberg AI, Autonomous Systems and Software Program (WASP) Joint Project; and in part by the Singapore NRF National Satellite of Excellence, Design Science and Technology for Secure Critical Infrastructure under Grant NSoE DeST-SCI2019-0012. This work was also supported in part by National Natural Science Foundation of China under Grants 62001419 and 62131016, in part by Zhejiang Provincial Natural Science Foundation of China under Grant LQ21F010012. This work is also funded by Dalian University of Technology, China under Grant No. DUT20RC(3)029.


\end{document}